\documentclass[twocolumn,prl,preprintnumbers,amsmath,amssymb,superscriptaddress]{revtex4}

\usepackage{amsmath}
\usepackage{amssymb}
\usepackage{bm}

\usepackage[dvips]{graphicx}
\usepackage{color}

\usepackage[dvipdfmx]{hyperref}

\allowdisplaybreaks
\begin{document}
\setlength{\oddsidemargin}{0cm}

\title{Thermal Emission from Semi-classical Dynamical Systems }

\author{Takeshi {\sc Morita}}
\email[]{morita.takeshi(at)shizuoka.ac.jp}
\affiliation{{\it Department of Physics, Shizuoka University, 836 Ohya, Suruga-ku, Shizuoka 422-8529, Japan 
}}
\affiliation{{\it 
		Graduate School of Science and Technology, Shizuoka University,
		836 Ohya, Suruga-ku, Shizuoka 422-8529, Japan
}}

\begin{abstract}
Recently the bound on the Lyapunov exponent  $\lambda_L \le 2\pi T/ \hbar$ in thermal quantum systems was conjectured by Maldacena, Shenker, and Stanford.
If we naively apply this bound to a system with a fixed Lyapunov exponent $\lambda_L$, it might predict the existence of the lower bound on temperature $T \ge \hbar  \lambda_L/ 2\pi $.
Particularly, it might mean that chaotic systems cannot be zero temperature quantum mechanically.
Even classical dynamical systems, which are deterministic, might exhibit thermal behaviors once we turn on quantum corrections.
We elaborate this possibility by investigating semi-classical particle motions near the hyperbolic fixed point and show that indeed quantum corrections may induce energy emission which obeys a Boltzmann distribution. 
We also argue that this emission is related to acoustic Hawking radiation in quantum fluid.
Besides, we discuss when the bound is saturated and show that a particle motion in an inverse harmonic potential and $c=1$ matrix model may saturate the bound although they are integrable.

\end{abstract}


\maketitle
\paragraph*{Introduction.---}
Understanding the quantum gravity is one of the most important problems in theoretical physics.
Through the developments in gauge/gravity correspondence \cite{Maldacena:1997re, Itzhaki:1998dd}, many people expect that large-$N$ gauge theories may illuminate the nature of the quantum gravity.
However only special classes of the large-$N$ gauge theories which possess certain properties may describe the gravity; understanding what the essential properties are in the gauge theories that have dual gravity is a crucial question.

Recently the idea of the maximal Lyapunov exponent was proposed  by Maldacena, Shenker, and Stanford, and it might capture one of essences of the gauge/gravity correspondence \cite{Maldacena:2015waa}.
They conjectured that thermal many-body quantum systems have an upper bound on the Lyapunov exponent: 
\begin{align}
\lambda_L \le \frac{2\pi T}{\hbar}  ,
\label{L-bound}
\end{align}
 where $T$ is temperature of the system and $\lambda_L$ is the Lyapunov exponent. 
 We have taken $k_B=1$.
 (See Ref.\,\cite{PhysRevE.98.012216} also.)
Particularly, if a field theory has the dual black hole geometry, the gravity calculation predicts that the field theory should saturate this bound $\lambda_L = 2\pi T/ \hbar$ \cite{Shenker:2013pqa, Shenker:2013yza, Shenker:2014cwa}.
(This would be related to the conjecture that the black hole may provide the fastest scrambler in nature \cite{Sekino:2008he}.)
This is called the maximal Lyapunov exponent, and the properties of this bound are actively being studied.
One remarkable example is the SYK model \cite{Sachdev:1992fk, Kitaev:2015}, which saturates the bound, and now people are exploring the dual gravity of this model.

Apart from the interest in the gauge/gravity correspondence, the possible existence of the bound on the Lyapunov exponent is interesting in its own right.
Especially, we can rewrite the bound Eq.\,\eqref{L-bound} as a more suggestive form
\begin{align}
T
 \ge \frac{\hbar}{2\pi }  \lambda_L.
\label{T-bound}
\end{align}
This relation tells us that the temperature of the chaotic system is bounded from below  \cite{Kurchan:2016nju}.
This is a striking prediction in semi-classical chaotic systems.
Suppose we consider a model of chaos with a finite Lyapunov exponent $\lambda_L$ in a classical Hamiltonian dynamical system, which is non-thermal and deterministic. (Indeed many of the studies of chaos have been developed in such a setup.)
Then the inequality Eq.\,\eqref{T-bound} is satisfied trivially as $T=0 \ge 0$.
Here temperature is zero because the system is non-thermal and the right-hand side is also zero because $\hbar=0$ in the classical model.
Now we consider the quantum corrections, and ask what will happen in the semi-classical regime.
Then the right-hand side of the inequality Eq.\,\eqref{T-bound} may become non-zero
\begin{align}
\frac{\hbar}{2\pi} \left(\lambda_L +O(\hbar) \right) =
\frac{\hbar}{2\pi} \lambda_L  +O(\hbar^2) ,
\end{align}
where we have assumed that the quantum corrections to the classical Lyapunov exponent $\lambda_L$ are $O(\hbar)$ \footnote{Although the definition of the Lyapunov exponent in quantum chaotic systems has not been established, the classical Lyapunov exponent would be well-defined semi-classically within the Ehrenfest's time.
} and they can be ignored at the leading order.
Thus, if the bound Eq.\,\eqref{T-bound} is correct, at least an $O(\hbar)$ temperature has to be induced in the system somehow.

Such a possibility of the emergence of thermal natures in non-thermal classical systems reminds us of the black hole thermodynamics.
Although black holes are just classical solutions of general relativity, they behave as the thermal baths through the Hawking radiation in the semi-classical regime \cite{Hawking:1974rv, Hawking:1974sw}.
Thus the bound Eq.\,\eqref{T-bound} might imply that Hawking radiation like phenomena are observed in semi-classical chaotic systems.

Note that the original proposal of the bound Eq.\,\eqref{L-bound} was discussed in thermal many-body quantum systems \cite{Maldacena:2015waa}.
On a related note, many of the studies of the bound have been investigated in the situations where the Lyapunov exponents depend on temperatures \cite{Shenker:2013pqa, Shenker:2013yza, Shenker:2014cwa}.
Hence it is unclear whether the bound Eq.\,\eqref{L-bound} really implies the emergence of thermal behaviors in the classically non-thermal chaotic systems whose (classical) Lyapunov exponents do not depend on temperatures.

The aim of this Letter is to pursue this possibility in Hamiltonian dynamical systems described by particle(s).
If the system is chaotic, then, typically, a hyperbolic fixed point exists, and we investigate particle motions near this fixed point.
We will see that indeed  energy emission that obeys a Boltzmann distribution is induced around the fixed point quantum mechanically.
We will also argue that this thermal emission is related to acoustic Hawking radiation \cite{Unruh:1980cg, Visser:1997ux,Giovanazzi:2004zv} if we prepare many particles and regard them as a quantum fluid. 
Besides, we discuss that the bound Eq.\,\eqref{T-bound} may be saturated even in integrable systems.

\paragraph*{Particle motions near hyperbolic fixed point.---}
We first consider particle motions in chaotic systems in classical mechanics.
Many chaotic behaviors arise through the two ingredients: hyperbolic fixed point and broken homoclinic orbit \cite{Wiggins:215055}.
The particle trajectories are stretched near the hyperbolic fixed point, and, through the broken homoclinic orbit, somehow the trajectories go back around the hyperbolic fixed point, and, by repeating these procedures, the trajectories develop complicated chaotic motions.
(We have in mind, for example, a driven pendulum motion.)

\begin{figure}
	\begin{center}
		\includegraphics[scale=1]{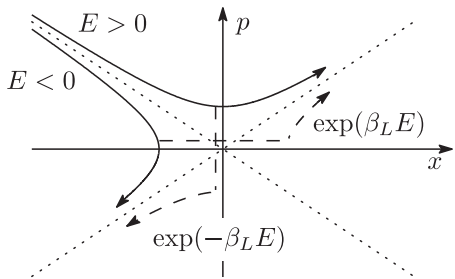}
		\vspace*{-2mm}
		\caption{The sketch of the trajectories of the incoming particles in the classical mechanics (solid lines) and the quantum corrections (broken lines) near the hyperbolic fixed point $(x,p)=(0,0)$ in the phase space.
			The dotted lines are the separatrices ($E=0$).
		}
		\label{Fig-hyper} 
		\vspace*{-5mm}
	\end{center}
\end{figure}

We will show that the hyperbolic fixed point plays the key role for the emergence of thermal behaviors in the quantum chaotic systems.
The particle motions near the hyperbolic fixed point might be effectively captured by the one-dimensional particle motions in an inverse harmonic potential
\begin{align}
m \ddot x(t) = -V'(x), \qquad V(x)=-\frac{\alpha }{2}x^2.
\label{inverse-EOM}
\end{align}
Here $m$ is the mass of the particle and $\alpha$ is the curvature of the potential.
The point $(x,p)=(0,0)$ in the phase space is the hyperbolic fixed point as a model for that of the chaotic system. (See Fig.\,\ref{Fig-hyper}.)
The solution of this equation is given by
\begin{align}
x(t)=c_1 e^{\sqrt{\alpha/m}t} + c_2 e^{-\sqrt{\alpha/m}t},
\label{particle-chaos}
\end{align}
where $c_1$ and $c_2$ are constants determined by the initial condition, and the sensitivity of the initial condition with the Lyapunov exponent is shown by
\begin{align}
\lambda_L=\sqrt{\frac{\alpha}{m}}.
\label{Lyapunov}
\end{align}
Although this Lyapunov exponent generally differs from that of the considered chaotic system,
if we naively apply the bound Eq.\,\eqref{T-bound}, we obtain the relation
\begin{align}
T \ge
T_L := \frac{\hbar}{2\pi} \sqrt{\frac{\alpha}{m}}.
\label{T-mini}
\end{align}
We will show that once we turn on the quantum corrections in the motion Eq.\,\eqref{particle-chaos}, thermal energy emission associated with the temperature $T_L$ is induced\footnote{Semi-classical thermal radiations from inverse harmonic potentials have been discussed in several contexts. See for example \cite{Brout:1995rd,Karczmarek:2004bw,Giovanazzi:2004zv,Betzios:2016yaq}.}.

\paragraph*{Thermal emission from hyperbolic fixed point.---}
Suppose a particle with energy $E$ moves toward the potential from the left ($x \to 0$).
Then, in classical mechanics, if the energy  is negative, the particle is reflected by the potential and goes back ($x \to - \infty $), while, if $E$ is positive, the particle goes through the potential toward $x \to + \infty$. 
(See Fig.\,\ref{Fig-hyper}.)

Here we consider the quantum corrections to this classical particle motion by solving the Schr\"{o}dinger equation with the Hamiltonian
\begin{align}
\hat H =  -\frac{\hbar^2}{2m} \frac{\partial^2}{\partial x^2} - \frac{\alpha}{2} x^2.
\end{align}
In the case of $E<0$, due to the quantum tunneling, the particle can penetrate the potential.
We can exactly compute this tunneling probability $P_T(E)$  by using the parabolic cylinder function \cite{Abramowitz} and obtain the following \cite{Moore:1991zv, Morita:2018sen}:
\begin{align}
 \quad P_T(E)= \frac{1}{\exp\left( -\frac{2\pi}{\hbar} \sqrt{\frac{m}{\alpha}} E \right)+1}
 = \frac{1}{\exp\left( - E/T_L \right)+1}
 .
\label{PT}
\end{align}
It means that the ratio of the probability of taking the classical trajectory ($x \to - \infty$) to that of the quantum one ($x \to + \infty$) is $1$ to $\exp\left( \beta_L E  \right)$ where $\beta_L:=1/T_L$ \eqref{T-mini}.

In the case of $E>0$, the incoming particle may be reflected by the potential quantum mechanically.
The probability of the reflectance of this process is given by
 \cite{Moore:1991zv, Morita:2018sen}
\begin{align}
P_R(E)= \frac{1}{\exp\left( \frac{2\pi}{\hbar} \sqrt{\frac{m}{\alpha}} E \right)+1}
= \frac{1}{\exp\left(  E/T_L \right)+1}.
\label{PR}
\end{align}
In this case, the ratio of the probability of taking the classical trajectory ($x \to + \infty$) to that of the quantum one ($x \to - \infty$) is $1$ to $\exp\left( -\beta_L E  \right)$.
(See Fig.\,\ref{Fig-hyper}.)

\begin{figure}
	\begin{center}
		\includegraphics[scale=1]{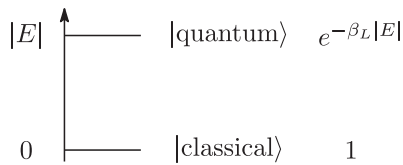}
		\vspace*{-2mm}
		\caption{The sketch of the relation between the two trajectories (classical and quantum) and the probability ratio. The ratio of the probability of taking the classical trajectory to that of the quantum one is $1$ to $\exp\left( - \beta_L |E|  \right)$, where $E$ is the energy of the particle.
			They can be regarded as the two level system and the quantum trajectory can be interpreted as an excited state.
		}
		\label{Fig-2-level} 
		\vspace*{-9mm}
	\end{center}
\end{figure}

In this way, the quantum corrections may change the classical trajectories of the particle motions to the new ones which are forbidden in classical mechanics.
The probabilities of taking the quantum trajectories in the two cases can be combined into the single form $\exp\left( - \beta_L |E|  \right)$, and the ratio is always given by 1 to $\exp\left( - \beta_L |E|  \right)$.
This ratio may be interpreted as the Boltzmann factor of the two level system at temperature $T_L$ where the ground state (zero energy) and the excited state (energy = $|E|$ ) correspond to the classical trajectory and quantum one, respectively.
See Fig.\,\ref{Fig-2-level}.

This interpretation of the two level system may be clarified by considering the energy transportation through these processes.
In the case of $E<0$, if the tunneling occurs, the negative energy particle is removed from the left region ($x<0$), and thus the energy in this region increases by $-E$($>0$) comparing with the classical process.
In the case of $E>0$, if the quantum reflection occurs, the particle carrying the positive energy comes into the left region, and again the energy in the left region increases by $E$.
Thus, in both cases, the quantum corrections induce the energy $|E|$ in the left region.
Hence the situation in the left region is really analogous to the two level system in which the emission of the energy $|E|$ occurs by the probability $\exp\left(-\beta_L |E| \right)$.

Therefore the particle motion near the hyperbolic fixed point $(x,p)=(0,0)$ shows the thermal behavior.
The fluctuations of the quantum mechanics imitate those of the thermodynamics.
Remarkably, the temperature saturates the bound Eq.\,\eqref{T-bound} predicted by Ref.\,\cite{Maldacena:2015waa}. (Of course, the Lyapunov exponent of \eqref{T-bound} differs from that of the whole system. We will discuss this point later.)
Note that this process may occur within the Ehrenfest's time $\sim \frac{1}{\lambda_L}\log\frac{1}{\hbar}  $ and the Lyapunov exponent $\lambda_L$ which has been obtained through the classical motion Eq.\,\eqref{particle-chaos} may be valid.
(Hence we may not need to employ the out-of-time-ordered correlator \cite{1969JETP...28.1200L} to evaluate the Lyapunov exponent.)
One surprise is that the chaotic natures of the system do not play any important role in this emission process.
Just the dynamics near the hyperbolic fixed point causes the emergence of the thermal properties.

\paragraph*{Connection to acoustic Hawking radiation.---}
So far we have seen that, if the particle with energy $E$ is injected to the inverse harmonic potential from the left, the energy flow occurs quantum mechanically and energy in the left region increases by $|E|$, which compares with the classical motion Eq.\,\eqref{particle-chaos} as if it is a thermal emission at temperature $T_L$.

\begin{figure}
	\begin{center}
		\includegraphics[scale=1]{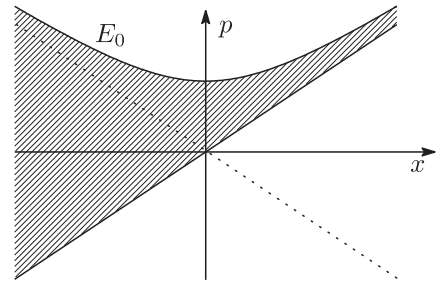}
		\vspace*{-2mm}
		\caption{The classical droplet of the Fermi fluid in the phase space.
			We consider the in-coming right moving Fermions up to energy level $E_0$.
		}
		\label{Fig-BH} 
		\vspace*{-5mm}
	\end{center}
\end{figure}

The emergence of thermal emission in such a semi-classical system reminds us of Hawking radiation.
Here we argue that this energy flow is indeed related to acoustic Hawking radiation in quantum fluid \cite{Unruh:1980cg, Visser:1997ux}.
(Related discussions have been done in Ref.\,\cite{Giovanazzi:2004zv}.)
Suppose that, instead of the single particle, we inject many free Fermions from the left toward the potential such that these right moving Fermions occupy the energy level up to $E_0 (>0)$\footnote{In chaotic systems, it would be hard to find the energy levels. 
	Here we consider just the inverse harmonic potential and implicitly introduce the IR cut off so that the Fermions are confined, and obtain the energy levels.}. (See Fig.\,\ref{Fig-BH}.)
Then the energy emission occurs at each energy level through the quantum effects and 
the energy density of the flux at location $x (<0)$ can be calculated as \cite{Morita:2018sen}
\begin{align}
	& \frac{m}{2\pi \hbar} \left( \int^0_{-\infty}   \frac{d E}{|p(E,x)|}
	\frac{-E }{e^{- \beta_L E }+1} +
	\int^{E_0}_{0}   \frac{d E}{|p(E,x)|}
	\frac{E }{e^{ \beta_L E }+1} \right) \nonumber \\
	&= \frac{1}{ (- x)} \frac{T_L}{24}\left(1 + O\left(\frac{T_L}{\alpha x^2}\right) \right).
	\label{energy-flow}
\end{align}
Here $p(E,x)$ is the classical momentum for the particle,
\begin{align}
	p(E,x)=\sqrt{2m\left(E+\frac{\alpha}{2}x^2\right)},
	\label{moment}
\end{align}
and it appears in Eq.\,\eqref{energy-flow} through the density of the Fermions in the phase space.
Here we have assumed that $E_0$ is sufficiently large.
This energy emission obviously obeys the Fermi-Dirac distribution, and, hence, it can be regarded as thermal.

On the other hand, since we are considering many Fermions, we can treat them as a one-dimensional Fermi fluid.
It is known that the Fermi fluid composed of the non-relativistic free Fermions classically obeys the following continuity equation and the Euler equation with pressure $p=\hbar^2 \pi^2 \rho^3/3 m^2$ \cite{Dhar:1992rs,Dhar:1992hr,Mandal:2013id},
\begin{align}
	&\partial_t \rho + \partial_x(\rho v)=0, \nonumber \\
	& \partial_t v + \partial_x \left(\frac{1}{2}  v^2  +  \frac{ \hbar^2 \pi^2}{2m^2} \rho^2 + \frac{1}{m}  V(x)\right)=0.
	\label{euler}
\end{align}
Here $\rho(x,t)$ and $v(x,t)$ denote the Fermion density  and velocity field, respectively, and $V(x)=- \alpha x^2/2$ in our case.  

By considering the classical fluid flow corresponding to Fig.\,\ref{Fig-BH}, we can show that the velocity of the flow exceeds the speed of the sound at $x=0$, and, if we turn on the quantum corrections, the acoustic Hawking radiation at temperature 
\begin{align}
	T= \frac{\hbar }{2\pi} \sqrt{\frac{\alpha}{m}}
	\label{Hawking-temp}
\end{align}
is emitted from $x=0$ to $x=-\infty$  \cite{Giovanazzi:2004zv, Morita:2018sen}.
Here the obtained temperature precisely agrees with the temperature $T_L$ Eq.\,(\ref{T-mini}) in our previous argument.
It is not difficult to reproduce the energy flow Eq.\,\eqref{energy-flow} as the emission through the acoustic Hawking radiation using the fluid mechanics \cite{Morita:2018sen}.
Therefore the energy emission near the hyperbolic fixed point discussed in the previous section can be regarded as the particle description of the acoustic Hawking radiation.

\paragraph*{Discussions.---}
We have argued for the mechanism of the emergence of the thermal emission in chaotic systems by investigating the particle motions near the hyperbolic fixed point through the inverse harmonic potential model Eq.\,\eqref{inverse-EOM}. 
If the particle with energy $E$ is injected to the inverse harmonic potential from the left, the energy flow occurs quantum mechanically, and the energy in the left region increases by $|E|$, which compares with the classical motion Eq.\,\eqref{particle-chaos} as if it is thermal emission at temperature $T_L$.

The idea that every chaotic system with finite Lyapunov exponents cannot be zero temperature quantum mechanically based on the bound Eq.\,\eqref{L-bound} sounds radical.
However, if a system has a hyperbolic fixed point and the effective one-dimensional description Eq.\,\eqref{inverse-EOM} can be applied, the thermal emission may be induced.
This might be analogous to the radiations from the black holes \cite{Hawking:1974rv, Hawking:1974sw} as we mentioned in the introduction.
This point has been emphasized in the example of the acoustic Hawking radiation in the Fermion system \cite{Giovanazzi:2004zv}.

One important question is about the temperature bound Eq.\,\eqref{T-bound}.
We have seen that the temperature of the radiation $T_L$ Eq.\,\eqref{T-mini} is fixed by the curvature of the potential $\alpha$ Eq.\,\eqref{inverse-EOM} and it saturates the temperature bound Eq.\,\eqref{T-bound}. 
However the Lyapunov exponent considered here is that of the hyperbolic fixed point and it differs from that of the whole system generally.
Thus it is unclear whether the bound Eq.\,\eqref{T-bound} is satisfied or not.
Naively the Lyapunov exponent of the system might be smaller than that of the hyperbolic fixed point, since the exponential time evolution of the particle motion Eq.\,\eqref{particle-chaos} would be disturbed  in the actual chaotic system, and the speed of the evolution might be decelerated.
If so, the bound Eq.\,\eqref{T-bound} is satisfied.
However, it may be not difficult to construct some models which violate the bound Eq.\,\eqref{T-bound}, although they might be artificial \cite{Kurchan:2016nju}.
It would be important to understand whether the bound on chaos Eq.\,\eqref{L-bound} works in the classically deterministic dynamical systems by developing our argument in this Letter.

Another important point is that the particle motion in the inverse harmonic potential Eq.\,\eqref{inverse-EOM} saturates the bound on the Lyapunov exponent Eq.\,\eqref{L-bound}.
Since the maximal Lyapunov $\lambda_L=2 \pi T/ \hbar $ was supposed to be related to the maximal chaos \cite{Maldacena:2015waa} or the fastest scrambler \cite{Sekino:2008he}, it is surprising that the free particle motion Eq.\,\eqref{particle-chaos} saturates this bound.
On the other hand, if we consider many free Fermions in the inverse harmonic potential, the system can be mapped to the two-dimensional string theory through the $c=1$ matrix model \cite{Klebanov:1991qa,Ginsparg:1993is,Polchinski:1994mb}.
Therefore the saturation of the bound might be related to the existence of the gravity description, although the free Fermions may not be able to describe any two dimensional black holes \cite{Karczmarek:2004bw}.
It would be interesting if we could reveal why the bound is saturated from the point of view of the two-dimensional string theory, and 
we leave this problem for future investigations.

Finally we discuss a possible connection to the original conjecture of the bound on chaos in quantum many-body systems \cite{Maldacena:2015waa}.
Suppose that there are $N$ interacting classical particles at temperature $T$.
Then the system possesses $2N$ Lyapunov exponents: $ \pm \lambda_1, \cdots,  \pm \lambda_N$, where they would depend on $T$ and we have assumed that the Hamiltonian is time reversal.
(We also assume that $\lambda_N$ is the maximum one.)
This system may have two time scales: the dissipation time  $t_d$  and  the scrambling time $t_*$  \cite{Maldacena:2015waa}, and we may observe the exponential developments of the deviations of the observables $\delta X$ between these two time scales $t_d \le t \ll t_*$.
Thus, in this time scale, there is a mode $\delta X_N$ that effectively feels the potential  $- \frac{m}{2} \lambda_N^2 \delta X_N^2$, causing the exponential development.
Then, through the mechanism of this Letter, this mode would be disturbed quantum mechanically as if it has the temperature $T_{\rm eff}\sim \frac{\hbar}{2\pi} \lambda_N $.
If $T \gg T_{\rm eff}$, this effect may be irrelevant .
However, if $T \ll T_{\rm eff}$, the quantum fluctuations of $\delta X_n$ may overcome the thermal fluctuations, and the thermal equilibrium state may be disturbed.
Therefore, such a large Lyapunov exponent $\lambda_N$ may be forbidden.
It may intuitively explain the bound on chaos.

The author would like to thank Koji Hashimoto, Pei-Ming Ho, Satoshi Iso, Shoichi Kawamoto, Manas Kulkarni, Gautam Mandal, Yoshinori Matsuo, Joseph Samuel, and Asato Tsuchiya for valuable discussions and comments.
The work of T.~M. is supported in part by Grant-in-Aid for Young Scientists B (No. 15K17643) from JSPS.

{\normalsize \bibliographystyle{JHEP} \bibliography{AHR} }

\end{document}